# Predicate Logic with Definitions

Victor Makarov

EMD Inc
186 Bay 31$^{st}$ Street
Brooklyn, NY 11214
USA

vedasystem@aol.com

## Abstract

Predicate Logic with Definitions (PLD or D-logic) is a modification of first-order logic intended mostly for practical formalization of mathematics. The main syntactic constructs of D-logic are terms, formulas and definitions. A definition is a definition of variables, a definition of constants, or a composite definition (D-logic has also abbreviation definitions called abbreviations). Definitions can be used inside terms and formulas. This possibility alleviates introducing new quantifier-like names. Composite definitions allow constructing new definitions from existing ones.

## 1. Introduction

The importance of practical formalization of mathematics has been widely recognized now – both for mathematics itself and especially for various applications such as, for instance, computer hardware and software design [Friedman 97, Qed Manifesto 94]. Following [Harrison 96], by *practical formalization of mathematics* "we mean expressing mathematics, both statements and proofs, in a (usually small and simple) formal language with strict rules of grammar and unambiguous semantics". Such a language can be naturally called as a *practical formal mathematical language* [Glushkov 72].
   One of the most well-known such languages is the Mizar language [Trybulec 93] (though the language is not very simple – a BNF description of the Mizar syntax is 10 pages long, there are about 140 nonterminals in the BNF grammar, and no complete description of the Mizar "semantics" has been yet published). The Mizar language is based on first-order logic (the language has also some not first-order features such as free second order variables used in axiom schemas). The Ontic language [McAllister 88] is another such language based on first-order logic. But most other such languages are based on higher order logic, e.g. [Gordon 93] (HOL).
   From the other side, the development of specification languages such as Z [Spivey 92], VDM SL [Jones 90], and modern object – oriented programming languages such as C++ [Stroustrup 97], Eifell [Meyer 92], Smalltalk [Kay 96] is very interesting for the design of practical logical languages . In this connection, see, e.g. [Tseytin 98], where the term "object-oriented logic" has been coined. In particular, Tseytin writes:
    "The fundamental concepts of object-oriented programming like *object*, *class*, i*nheritance* … belong to description of thinking rather than to programming techniques".

Because of their importance, let us discuss in some details the main concepts of object-oriented programming (OOP). The main concepts of OOP are the concepts of *object* and *class*. An object is a k-tuple of *values* (k ≥ 1). A value can be, for example, an integer or real number, a character or a string of characters or another object. Any object is an *instance* of a class. A class is a description (in an *object-oriented programming language*) of the structure of its objects, some conditions (called *class invariant* in the Eifell language) that all objects of the class must satisfy and some operations on its objects. A class is to an object (of its class) as a blueprint of a real object (a bicycle, for example) is to the real object.

Note the similarity between the concept of object in OOP and the concept of mathematical object. A mathematical object, such as a group, a graph and so on, is also usually defined as a k-tuple of sets, functions or other mathematical objects. For example, a group is a pair( or a 2-tuple) <G, f> where G is the carrier set and f is a function from G×G to G, satisfying certain conditions – group axioms. Note that a concrete group is a *model* of the group theory and, more generally, a concrete mathematical object is a model of the corresponding theory (excluding *primitive* mathematical objects such as, for example, integer numbers). So we may say that a class is to an object (of the class) as a mathematical theory is to a model of the theory (and the class invariant is the analogue of the axioms of a theory).

As mathematical theories are the main way of structuring of the mathematical knowledge, so classes are the main way of structuring of the "programming knowledge". A very important feature of OOP is the possibility to build class hierarchies by defining a class as a heir of other class (simple inheritance) or classes (multiple inheritance). In mathematics, a similar method is used when we define, for example, the theory of linear ordered groups referring to the theory of linear order and the group theory.

Unfortunately, classical logical languages (such as first-order logic) have no special syntactic constructs for definitions of mathematical theories (actually, they have no syntactic constructs for definitions at all). It is the main obstacle in using a classical logical language as a practical formal mathematical language.

In this paper, Predicate Logic with Definitions (PLD or D-logic) intended mostly for practical formalization of mathematics is introduced. As Tseytin's "object-oriented logic", D-logic has been also developed under influence of object-oriented programming. But, unlike Tseytin's "object-oriented logic" (which is not based on the standard (i.e. first-order) logic), the D – logic is a modification of first-order logic by adding to the main syntactic constructs of first-order logic (terms and formulas) new syntactic constructs called *definitions*. A definition is a definition of variables, a definition of constants, or a composite definition. Definitions of constants in D-logic are in a sense analogous to classes in object-oriented programming languages and they are mostly used for defining mathematical theories. Definitions of variables are mostly used in quantified formulas and terms. Such an approach (using definitions inside formulas and terms) alleviates introducing new quantifier-like names. Composite definitions allow constructing new definitions from existing ones. It corresponds (in the case of definitions of constants) to defining a new class in object-oriented programming as a heir of early defined classes.

D-logic is intended to be used as the input language of a new version of the Veda proof-checking system (main ideas of the Veda project have been described in [Makarov 92]).

## 2. From first-order logic to D – logic

The symbols used in the language of first-order logic (we follow basically to [Davis 93] and [Shoenfield 73]) are usually divided into the following classes:

1. Logical symbols: ~ (negation), → (implication), & (conjunction), ∨ (disjunction), ≡ (equivalence), ∀ (universal quantifier), ∃ (existential quantifier).
2. Constant symbols.
3. Function symbols.
4. Relation symbols.

For each function and relation symbol, a natural number k, called *arity*, must be assigned. The notions of terms and formulas can then be defined [Davis 93]. Fixing the sets of constant, function and relation symbols we receive a concrete first-order language – for instance, the language of the ZFC set theory.

Let us consider the formulas of the form ∃xP and ∀xP, where x is a variable, P is a formula. Let us slightly change the syntax of such formulas and write ∃(x|P) and ∀(x|P), respectively, where '|' is a punctuation symbol. Let us call (in this section) the expression x|P as a *definition of variables* (actually, a definition of variables in D-logic can contain many variables – see below).

An immediate benefit of this approach is the following: it is easier to introduce new "quantifier-like" names. To show it, let us generalize the notion of arity in the following way. Let us call a *generalized arity* (or g-arity or simply arity) any finite sequence of the letters F, T, D (from Formula, Term, Definition). We assign a g-arity to each symbol as follows:

1.1 Unary logical symbols (~): FF.

1.2 Binary logical symbols (→, &, ∨, ≡ ): FFF.

1.3 Quantifiers (∀,∃): DF.

2. Constant and variable symbols: T.

3. Function symbols of arity k: $T^{k+1}$ (i.e. the sequence of k+1 letters T).

4. Relation symbols of arity k: $T^k F$. For instance, the g-arity TTF will be assigned to the symbols "=" (equality) and "∈" (is a member of).

Now it is obvious that for introducing, for example, Hilbert's ε-symbol or the symbol λ used in λ-notation, it is sufficient to assign to the symbols ε and λ the g-arities DT and DTT respectively (and, of course, provide defining axioms for the symbols). So the expressions εxP(x) and λxf(x) could be written in the following way: ε(x|P(x)) and λ(x|true, f(x)) respectively. But actually, due to practical considerations, a slightly

different syntax is used: H[x|P(x)] and F[x|true, f(x)] respectively (typed λ-terms of the form λx:t f(x) can be written as F[x|x∈t, f(x)] or simply as F[x:t, f(x)] ).

Besides of definitions of variables, D-logic has also *definitions of constants* which have (in the simplest case) the form def[$c_1$, ..., $c_k$ | P] (i.e. they have the same abstract syntax as the definitions of variables) where k > 0, $c_1$, ..., $c_k$ are some constant symbols, P is the defining axiom of the constant symbols.

Note that it is possible to introduce symbols of the g-arities, for example, DD, DDD or DFD. Such symbols can be used for denoting *operations on definitions* and constructing *composite definitions*.

For example, if d is a definition of variables of the form (x|P), R is a formula then the negation of the definition d (symbolically, ~d) is the definition (x|~P). And the conjunction of the definition d and the formula R (symbolically, d&R) is the definition (x|P&R). Another useful operations on definitions are *concatenation* and *hereditary concatenation* of two definitions (the names "!" and "%" will be used for these operations, respectively).

If d1 is ($x_1$, ... ,$x_k$ | P), d2 is ($y_1$, ... ,$y_m$ | Q)  then:
     d1 !  d2 is ($x_1$, ... ,$x_k$, $y_1$, ... ,$y_m$ | P & Q),
     d1 %  d2  is ($x_1$, ... ,$x_k$, $y_2$, ... ,$y_m$ P & Q'),
where Q' is the result of substitution, in the formula Q, the variable $x_1$ instead of the variable $y_1$. Though definitions of variables are used in the examples above, the same operations are also can be used for definitions of constants (because both definitions of constants and definitions of variables have the same abstract syntax).

It is widely accepted now that every mathematical theory T can be considered as an extension of the ZFC set theory by adding to ZFC (or an extension of ZFC) the constants and axioms of the theory T [Dieudonne 82, p.215 ]. So definitions of constants can be used for defining theories.

For example a definition of group can have the following form:
     Group  := def[G, * ; <group axioms>]
where the constants G and * represent respectively the carrier set and the law of composition (a function from G x G to G ), the symbol ":=" means that the name Group is an abbreviation of the expression "def[G, * ; <group axioms>]" – see below a description of abbreviations. Note that the formula  E[Group] expresses the following statement: the group theory is consistent.

Using the & operation on definitions, a definition of commutative group can be written in the following form:
CommutativeGroup := Group & A[x, y | x ∈ G & y ∈ G → x * y = y * x].
Using the definition of group Group and a definition of linear order (let us name it as LinearOrder), a definition of linear ordered group can be written in the following way:
     LinearOrderedGroup := Group % LinearOrder & R
where % is the operation of hereditary concatenation (see above) and R is an additional axiom.

For denoting symbols we shall use both *internal* and *external* names. An internal name can be identified with the symbol itself. All internal names must be unique. Because it is inconvenient to use internal names, external (usually more mnemonic but not necessary unique) names are used. It is always must be possible for every external name to determine its corresponding internal name (as a rule, using the context). In a computer implementation of D–logic, internal names can be chosen automatically, when the system handles definitions. Usually, an external name is an identifier, an operator or a bracketed name. An identifier is a sequence of letters or digits beginning from a letter. Usually, an operator is a sequence of such characters as +, –, *, /, %, &, ~, #, =, ^. Mostly, operators are used for constructing infix expressions. A bracketed name is an identifier after which a left bracket follows. Internal names can be considered as related to abstract syntax [Sethi 96], whereas external names are related to concrete syntax.

## 3. D-languages

To specify a (logical) language L we must specify the set of its symbols and the set of formulas of L (the language L can also have other syntactic constructs such as e.g. terms). The main syntactic constructs of a D-language are formulas, terms and definitions which are called *basic constructs* (it is convenient to include into a D-language some auxiliary syntactic constructs called *declarations* – see below). For each basic construct, we assign a mode – one of the letters T, F, D in the following way:
  Terms are assigned the mode T, formulas are assigned the mode F, and definitions are assigned the mode D.
   A D-language L has as symbols the following:

a) *Variables* – an infinite set V;
b) *Primary symbols* – a nonempty set S, such that $S \cap V = \emptyset$. For each primary symbol s, a g-arity g(s) must be assigned – see section 2 (from now on we shall write simply "arity" instead of "g-arity"). The set S must include the set of *logical symbols LS* (the arities of the logical symbols are shown on the right): ~ (negation) FF, & (conjunction) FFF, ∨ (disjunction) FFF, → (implication) FFF, ≡ (equivalence) FFF, if (if-then-else) FTTT, true (a truth value) F, false (a truth value) F, ∀ (the universal quantifier) DF, ∃ (the existential quantifier) DF, H (Hilbert's epsilon symbol) DT, = (equality) TTF. The symbols A, E may be used instead of the symbols ∀, ∃ due to practical considerations. The other (i.e. not logical) symbols are called *nonlogical symbols*. We assume that every D-language has the symbol ∈ with the arity TTF and the symbol → of the arity TTT among its nonlogical symbols (A→B will denote the set of all functions from A to B).
c) *Punctuation symbols* – the set P of such symbols as parenthesis ('(',')'), brackets ('[', ']'), comma (','), semicolon (';'), vertical bar ('|') and some other symbols (see below).

Let us call the quadruple U = (V, S, P, g) as a *vocabulary* (of the D-language). An *expression* in the vocabulary U is a finite sequence of symbols from the set $V \cup S \cup P$. Given the vocabulary U, we define the following sets (below last(g(s)) denotes the last element of the sequence g(s) ):
C = {s:S | g(s) = T} – the set of *constant symbols* (or simply *constants*);
R = {s:S | last(g(s)) = 'F'} – the set of *relation symbols*;
F = {s:S | last(g(s)) = 'T'} – the set of *function symbols*;
D = {s:S | last(g(s)) = 'D'} – the set of *definition symbols*.

## 3.1 Declarations

*Declarations* are auxiliary syntactic constructs (actually, they should be considered as metalanguage constructs). As definitions, declarations are used for introducing new names. But, unlike definitions, declarations can not be used inside terms, formulas or definitions. A declaration can be an abbreviation, a declaration of primary names, a declaration of variables, a declaration of an axiom or a declaration of a theorem.

*An abbreviation* has the form N := B, where N is an identifier (i.e. a sequence of letters and digits beginning from a letter), B is a term or a formula or a definition. The purpose of abbreviations is obvious: to use a possibly short name N instead of possibly large terms, formulas or definitions. The name N can not occur (directly or indirectly) in B.

If B is a formula then the abbreviation is called *a formula abbreviation*.

*A declaration of primary names* can have one of the following forms:

$$\text{def}[N : A] \quad \text{or}$$
$$\text{def}[N : A; u_1 = u_2] \quad \text{or}$$
$$\text{def}[N : A; p_1 \equiv p_2]$$

where N is a primary name (i.e. the name denoting a primary symbol) – an identifier or operator or a bracketed name, A is its arity, $u_1$ is a term of the form N or of the form $N(v_1, \ldots, v_k)$ where $v_1, \ldots, v_k$ (k > 0) are some variables (including propositional and definitional ones), $u_2$ is a term, $p_1$ is a formula of the form $N(v_1, \ldots, v_k)$, $p_2$ is a formula. All free variables of $u_2$ (or $p_2$) must be among the variables $v_1, \ldots, v_k$. The formula $u_1 = u_2$ (or $p_1 \equiv p_2$) is called the defining axiom of the name N. A declaration of primary names can be understood as a command to extend the current theory T by adding to the primary names of the theory T the new primary name N and the corresponding defining axiom. The form of the defining axioms guarantees that the extension of the theory T will be a conservative [Shoenfield 73] one.

*A declaration of variables* is used for introducing *syntactic variables* [Shoenfield 73] and has the form $V_1, \ldots, V_k : A$ where $V_1, \ldots, V_k$ are syntactic variable names (some identifiers), A is an one-letter arity (i.e. A is T, F, or D). Syntactic variables are used for writing axiom schemas. Syntactic variables of the arity T are called term variables, of the arity F – propositional variables and of the arity D – definitional variables.

*A declaration of an axiom* has the form Axiom A, where A is a formula abbreviation of the form N := Q. The declaration Axiom N := Q states that the formula Q is an axiom (and N is its name).

*A declaration of a theorem* has the form Theorem T, where T is a formula abbreviation of the form N := Q. The declaration Theorem N := Q states that the formula Q is possibly a theorem (and N is its name). A computer implementation of D-logic may try to find a proof of the formula Q.

## 3.2 Definitions

*A definition* can be a simple definition or a composite definition. *A simple definition* is a definition of constants or a definition of variables.

*A definition of constants* can be typed or untyped. *An untyped definition of constants* has the following form:

(1) $\quad\quad\quad \text{def}[c_1, \ldots, c_k; P]$

where $c_1, \ldots, c_k$ ($k > 0$) are constant names (identifiers or operators), P is a formula called the defining formula of the definition (the defining formula can not be considered as an axiom, because, e.g. P may be the formula "false"). If P is the formula "true", then the definition (1) can be written as def[$c_1, \ldots, c_k$].

*A typed definition of constants* has the following form:

(2)  def[$c_1:t_1, \ldots, c_k:t_k$; P]

where $c_1, \ldots, c_k$ ($k > 0$) and P have the same meaning as above and $t_1, \ldots, t_k$ are *types*. Each type is a term or a definition (if P is the formula "true", then it can be omitted).
. A typed definition of the form (2) corresponds to the following untyped definition of constants:

(3)  def[$c_1, \ldots, c_k$; $c_1 \in t_1$ & $\ldots$ & $c_k \in t_k$ & P]

*A **definition of variables*** can be typed or untyped or short. *An untyped definition of variables* has the following form:

(4)  dev[$x_1, \ldots, x_k$; P]

where $x_1, \ldots, x_k$ ($k > 0$) are variable names (some identifiers), P is a formula called the defining formula of the definition (again, the defining formula can not be considered as an axiom). If P is the formula "true", then the definition (4) can be written as dev[$x_1, \ldots, x_k$].

*A typed definition of variables* has the following form:

(5)  def[$x_1:t_1, \ldots, x_k:t_k$; P]

where $x_1, \ldots, x_k$ ($k > 0$) and P have the same meaning as above and $t_1, \ldots, t_k$ are types (if P is the formula "true", then it can be omitted). A typed definition of the form (5) corresponds to the following untyped definition of constants:

(6)  def[$c_1, \ldots, c_k$; $c_1 \in t_1$ & $\ldots$ & $c_k \in t_k$ & P]

*A short definition of variables* has the form x:t where x is a variable name (an identifier) and t is a type. The definition x:t denotes the definition dev[x:t].

*A composite definition* has the form $g(z_1, \ldots, z_k)$, where g is a definition symbol of the arity $m_1\ldots m_k D$ ($k > 0$), and $z_1, \ldots, z_k$ are basic constructs of the modes $m_1, \ldots, m_k$ respectively. If g is an infix name (or unary prefix name) then the notation $z_1$ g $z_2$ (or respectively g $z_1$) must be used. We shall assume that for each definition symbol is given a computable function that converts the composite definition into a simple one (i.e. definition symbols are eliminable).

It will be also assumed that there is a predefined definition symbol Rep of the arity DTDD. If d, d1 are definitions, $z = [z_1, \ldots, z_k]$ is a model of d (for a definition of model in D-logic, see below section 6), then Rep(d, z, d1) denotes the result of substitution in the definition d1 the terms $z_1, \ldots, z_k$ instead of the corresponding d-names in d1. Rep(d, z, d1) can be written as d(z, d1).

## 3.3 Terms

(1) Every variable introduced in a definition of variables is a term.

(2) Every primary name of the arity T is a term.

(3) Every constant introduced in a definition of constants is a term.

(4) If f is a function symbol of the arity $m_1\ldots m_kT$ (k > 0), and $z_1, \ldots, z_k$ are basic constructs of the modes $m_1, \ldots, m_k$ respectively then $f(z_1, \ldots, z_k)$ is a term. If f is an infix name (or an unary prefix name or a bracketed name "h[" - such names are called nonstandard) then the notation $z_1$ f $z_2$ (or respectively f $z_1$ or h[$z_1, \ldots, z_k$]) must be used.

(5) If d is a definition, z and x are terms, then d(z, x) is a term. Usually, $z = [z_1, \ldots, z_k]$ is a model of d; d(z, x) is the result of substitution the terms $z_1, \ldots, z_k$ instead of the corresponding d-names in x ; in other words, d(z, x) is "the value of x in the model z").

(6) If z, x are terms, then z.x is a term. Usually, z is a model of a definition d and the definition d can be found from the context. Then z.x denotes the term d(z, x).

(7) If $z_1, \ldots, z_k$ are terms (k ≥ 0) then { $z_1, \ldots, z_k$ } is a term ("the set consisting of $z_1, \ldots, z_k$").

(8) If $z_1, \ldots, z_k$ are terms (k ≥ 0) then [ $z_1, \ldots, z_k$ ] is a term ("the tuple ($z_1, \ldots, z_k$)").

(9) if d is a definition then [d] is a term. Suppose d has the form ($x_1, \ldots, x_k$|P). Then if k = 1 then [d] denotes $x_1$, otherwise [d] denotes the tuple [$x_1, \ldots, x_k$].

(10)   There are no other terms.

## 3.4 Formulas

(1)   Every primary name of the arity F (i.e. true and false) is a formula.

(2)   If r is a relation symbol of the arity $m_1\ldots m_kF$ (k > 0), and $z_1, \ldots, z_k$ are basic constructs of the modes $m_1, \ldots, m_k$ respectively then $r(z_1, \ldots, z_k)$ is a formula. If r is a nonstandard name then the corresponding notation must be used.

(3)   If d is a definition, z is a term, and p is a formula, then d(z, p) is a formula. Usually, $z = [z_1, \ldots, z_k]$ is a model of d; d(z, p) is the result of substitution the terms $z_1, \ldots, z_k$ instead of the corresponding d-names in p ; in other words, d(z, p) is "the value of p in the model z".

(4)   If d is a definition, z is a term, then d(z) is a formula (" z is a model of d"). If d is (x|P(x)) then d(z) denotes the formula P(z).

(5)   There are no other formulas.

## 4. D – theories

A D-theory T is a triple <L, A, R> where L is a D-language, A is a set of *axioms* (every axiom is a formula of L), R is a set of *inference rules* (or simply rules). Each inference rule is a computable function from the set Form(L)$^n$ to the set Form(L) where n is a natural number given for each inference rule.

The set A (of axioms) is the union of two sets, $A_L$ and $A_N$ – the sets of *logical* and *nonlogical* axioms, respectively. The logical axioms of a D-theory are the following formulas (below d is a definitional variable, z is a term variable, p is a propositional variable; note that if an axiom contains syntactic variables then it actually represents an axiom schema and denotes an infinite set of axiom):
1) Tautologies: every tautology is an axiom;
2) Each formula of the form d(z) → d(H[d]) is an axiom;     (ε-axiom)
3) Each formula of the form E[d] ≡ d(H[d]) is an axiom;     (the defining axiom for the existential quantifier)
4) Each formula of the form A[d] ≡ d(H[~d]) is an axiom;    (the defining axiom for the universal quantifier)
5) Each formula of the form E[d, p] ≡ E[d & p] is an axiom;    (the defining axiom for the bounded existential quantifier)
6) Each formula of the form A[d, p] ≡ A[d → p]) is an axiom;    (the defining axiom for the bounded universal quantifier)

The set R (of inference rules) consists of the following rule:
1) Modus ponens (MP): from the formulas P and P → Q infer the formula Q.

The set of *theorems* of the D-theory T can be defined in the usual way:
1) The axioms of T are theorems of T;
2) If all of the hypotheses of a rule of T are theorems of T then the conclusion of the rule is a theorem of T;
3) Any theorem of T may be obtained only using these definitions.

A D-theory T is called *consistent* if the formula "false" is not a theorem of T.

A *D-calculus* is a D-theory without nonlogical axioms.

Theorem. Any D-calculus is a consistent D-theory.
Proof: in a similar way as for first-order logic [Mendelson 63].

## 5. Some theorems of D-logic

Most of theorems in first-order logic have natural analogues in D-logic. In this section, such analogues of some theorems from [Hilbert 68, chapter 4, §3] will be listed (mostly without proofs, with the same numeration). The original theorems (maybe in a slightly different notation) will be written on the right.

a) A[d] → d(z);                    ( ∀xP(x) → P(z) )
b) d(z) → E[d];                    ( P(z) → ∃xP(x) )
   Proof: immediately from the ε-axiom d(z) → d(H[d]) and the defining axiom
        E[d] ≡ d(H[d]);
1. A[d] → E[d];                    ( ∀xP(x) → ∃xP(x) )
2. ~A[d] ≡ E[~d];                  ( ~∀xP(x) ≡ ∃x ~P(x) )
2.a ~A[d, P] ≡ E[d, ~P]
3`. ~E[d] ≡ A[~d];                 ( ~∃x P(x) ≡ ∀x ~P(x) )

3.a  ~E[d, P] ≡ A[d, ~P]
4.   A[P → d] ≡ P → A[d] (if P is d-free);  (∀x (P → Q(x)) ≡ P → ∀xQ(x))
1    A[P ∨ d] ≡ P ∨ A[d]; (if P is d-free);  (∀x (P ∨ Q(x)) ≡ P ∨ ∀xQ(x))
2    A[P & d] ≡ P & A[d]; (if P is d-free);  (∀x (P & Q(x)) ≡ P & ∀xQ(x))
3    A[d, P & Q] ≡ A[d, P] & A[d, Q];        (∀x (P(x) & Q(x)) ≡ ∀xP(x) & ∀xQ(x))
4    E[P ∨ d] ≡ P ∨ E[d]; (if P is d-free);  (∃x (P(x) ∨ Q(x)) ≡ P ∨ ∃xQ(x))
5    E[d, P ∨ Q] ≡ E[d, P] ∨ E[d, Q];        (∃x (P(x) ∨ Q(x)) ≡ ∃xP(x) ∨ ∃xQ(x))
6    E[P & d] ≡ P & E[d]; (if P is d-free);  (∃x (P & Q(x)) ≡ P & ∃xQ(x) )
7    A[d, P → Q] → (A[d, P] → A[d, Q]);      (∀x (P(x) → Q(x)) → (∀x P(x) → ∀x Q(x)) )
8    A[d, P → Q] → (E[d, P] → E[d, Q]);      (∀x (P(x) → Q(x)) → (∃x P(x) → ∃x Q(x)) )
9    A[d1, A[d2, P]] ≡ A[d2, A[d1, P]];      (∀x∀y P(x, y) ≡ ∀y∀x P(y, x))
10   (no natural analogue);                  (∀x∀y P(x, y) → ∀x P(x, x))
11   A[d1, A[d2, P & Q]] ≡ A[d1, P] & A[d2, P]; (∀x∀y (P(x) & Q(y)) ≡ ∀x P(x) &
                                                                    ∀y Q(y))
16a. A[d, P → Q] ≡ E[d, P] → Q; (Q is d-free); (∀x(P(x) → Q) ≡ ∃xP(x) → Q)
16b. E[d, P → Q] ≡ A[d, P] → Q; (Q is d-free); (∃x(P(x) → Q) ≡ (∀xP(x) → Q) )
17.  A[d, P ≡ Q] → (A[d, P] ≡ A[d, Q]);       (∀x (P(x) ≡ Q(x) → (∀xP(x) ≡ ∀xQ(x)))
18.  E[d1, A[d2, P]] → A[d2, E[d1, P]];       (∃x∀y P(x, y) → ∀y∃x P(x, y) )

The following formulas are also theorems:

21. (d → P)(z) ≡ d(z) → d(P, z)
22. (d & P)(z) ≡ d(z) & d(P, z)
23. (d ∨ P)(z) ≡ d(z) ∨ d(P, z)
24. (~d)(z) ≡ ~ (d(z))
25. A[d, P] & E[d] → P
25a. A[d, P] & d(z) → P
26. A[d, false] ≡ ~E[d]          (from 16.a, taking P = true, Q = false)
27. A[d, P ∨ Q] ≡ A[d&~P, Q]
28. A[d, P→Q] ≡ A[d& P, Q]

Some derived inference rules of D-logic are as follows:
1) from P → d(z) infer P → A[d], if P does not contain free occurrences of z (α-rule in [Hilbert 68] );
   Proof: substitute z = H[~d] in the formula "P → d(z)" and then use the defining axiom "A[d] ≡ d(H[~d])";
2) from d(z) → P infer E[d] → P, if P does not contain free occurrences of z (β-rule in [Hilbert 68] );
   Proof: substitute z = H[d] in the formula "P → d(z)" and then use the defining axiom "E[d] ≡ d(H[d])";
3) from d(z) infer A[d] ( γ` - rule in [Hilbert 68]);
   Proof: substitute z = H[~d] in the formula "d(z)" and the use the defining axiom "A[d] ≡ d(H[~d])";
4) from P infer A[d, P];
   Proof:

1) P;                    – given;
2) d(z) → P;             – from 1, we can choose z such that z does not occur in P;
3) d(z) → d(P, z) ;      – from 2, because d(P, z) = P
4) (d→P)(z);             – from 3 and Theorem 21;
5) A[d→P];               – from 4 by the inference rule 3 (γ`) above.
6) A[d, P];              – from 5 and the defining axiom A[d, P] ≡ A[d→P]

## 6. An important inference rule

We may assume that every mathematical theory T is an extension of the ZFC set theory by adding some constants $c_1, …, c_k$ and their defining axioms $a_1, a_2, …$ [Dieudonne 82].

A k-tuple $M = <z_1, … , z_k>$ where $z_i$ are some sets, is a *model* of T iff each formula Subst(a, $(c_1, …, c_k)$, $(z_1, … , z_k)$ ) where a is an axiom $a_1, a_2, …$ , is a theorem (the expression Subst(a, $(c_1, …, c_k)$, $(z_1, … , z_k)$ ) denotes the result of substitution, in the expression a, the terms $z_1, … , z_k$ instead of the constants $c_1, …, c_k$ , respectively.

Let P be a theorem (of theory T). Then the formula Subst(P, $(c_1, …, c_k)$, $(z_1, … , z_k)$ ) will be also a theorem. (see in this connection the "little theory version of axiomatic method" in [Farmer 92]).

In D-logic, this inference rule can be expressed in the following way: if a formula P is a theorem, a formula of the form d(z) is a theorem, then the formula d(P, z) is also a theorem − the definition d corresponds to the theory T, the formula d(z) can be understood as "z is a model of d", and the formula d(P, z) can be written as Subst(P, $(c_1, …, c_k)$, $(z_1, … , z_k)$ ).

Proof.
1. P                               (given)
2. d(z)                            (given)
3. A[d] → d(z)                     ( axiom a of D-logic)
4. A[d→P] → (d→ P)(z)              (from 3, take d→P instead of d)
5. A[d, P] →  (d(z) → d(P, z))     (from 4, A[d, P] ≡ A[d→P] and the formula 21)
6. A[d, P] & d(z) → d(P, z)        (from 5 and propositional calculus)
7. A[d, P]                         (from 1 and the rule 4- see the end of the previous section)
8. A[d, P] & d(z)                  (from 7, 2 and propositional calculus)
9. d(P, z)                         (from 6, 8 by MP)
End of proof

## 7. ZFC set theory

In this section, the ZFC set theory (we follow [Fraenkel 84]) will be introduced as a D-theory. The names a, b, f, s, t, u, x, y, z, A, B will be used as variables, the name d is a syntactic (definitional) variable, "//" means "comment".

def[∈ : "TTF"];                             // a member of
def[∉ : "TTF"; x ∉ y ≡ ~(x ∈ y) ];          // x is not a member of y
def[⊆: "TTF" ; a ⊆ b ≡ A[x:a, x ∈ b] ];     // a is a subset of b

Axiom Extensionality := a ⊆ b & b ⊆ a → a = b;

Axiom Pairing := A[x,y| E[z| A[u| u ∈ z ≡ u = x ∨ u = y]]];
def[ { : "TTT"; {x,y} = H[z| A[u| u ∈ z ≡ u =x ∨ u = y]]];   // H is Hilbert's epsilon-symbol
Theorem Pair := a ∈ {x,y} ≡ a = x or a = y;
def[ { : "TT"; {x} = {x,x} ];
Theorem Single := a in {x} ≡ a = x;

Axiom Union := A[a| E[y| A[x| x ∈ y ≡ E[z| x ∈ z & z ∈ a ]]]];
def[U[ : "TT"; U[a] = H[y| A[x| x ∈ y ≡ E[z| x ∈ z & z ∈ a ]]]]; // union
Theorem UnionTrm := x ∈ U[a] ≡ E[z| x ∈ z & z ∈ a ];
def[∪ : "TTT"; A∪B = U[{A,B}] ];            // A∪B is the union of A and B
Theorem TrmUnion := x ∈ A∪B == x ∈ A ∨ x ∈ B;

Axiom PowerSet := A[a| E[y| A[x| x ∈ y ≡ x ⊆ y ]]];
def[P : "TT"; P(a) = H[y| A[x| x ∈ y ≡ x ⊆ y ]];
Theorem PowerSetTrm := x ∈ P(a) ≡ x ⊆ a;

def[Set: "DF"; Set(d) ≡ E[z| A[x| x ∈ z ≡ d(x) ]]];

def[TypedSetDef: "DF"];   //Built-in : Typed definition where types are sets

Axiom Separation := TypedSetDef(d) → Set(d);  // actually, it is an axiom schema
def[ { : "DT"; {d} = H[z| A[x| x ∈ z ≡ d(x) ]]];
Theorem SetDefTrm := Set(d) → (x ∈ {d} ≡ d(x));

def[∩ : "TTT"; A∩B = {x:A & x ∈ B} ];  // A∩B is the intersection of A and B
Theorem Intersection := x ∈ A∩B ≡ x ∈ A & x ∈ B;

def[∩ : "TT"; ∩A = if(A = {}, {}, {x:H[y| y ∈ A] & A[z:A, x ∈ z})];
                // ∩A is the intersection of all sets in A
Theorem Intersection1 := A = {} → ∩A = {};
Theorem Intersection2 := A ≠ {} → ( x ∈ ∩A ≡ A[z:A, x ∈ z] );

set := def[anyset; true];

def[{} : "T"; {} = {x : anyset | false } ];   // {} is the empty set
Theorem EmptySet := ~(x ∈ {});

Axiom Infinity := E[A| {} in A & A[x:A, x ∪ {x} in A]];

def[RepAxDef : "DF"]; // Built-in : definition of the form y|E[x:A & P(x,y)]
                // where P(x,y) is a functional condition [Fraenkel 84] on A
// In particular, every formula of the form y = f(x) is a functional condition

Axiom Replacement := RepAxDef(d) → Set(d);   // axiom schema
def[{ : "DTT"; {d,f} = {y| E[x:{d} & y = d(f,x)]}];
Theorem ReplacementTrm := Set(d) → (y ∈ {d,f} ≡ E[x:{d} & y = d(f,x)]);

def[Singleton : "TF"; Singleton(A) ≡ E[x| A = {x}]];
def[× : "TT"; ×t = {x: P(U[t]) & A[s:t, Singleton(s ∩ x)]}];   // Outer Product
def[× : "TTT"; A×B = {x:A!y:B};                                // Cartesian Product
def[Disjoint : "TF"; Disjoint(A) == A[x:A!y:A, x ≠ y → x ∩ y = {}]];

Axiom Choice := Disjoint(A) & {} ∉ A → ×A ≠ {};

Axiom Foundation := y ≠ {} → E[u:y, u ∩ y = {}];

## 8. Relations and functions

In this section, the concepts of relation and function (and also some related concepts) are introduced (following [Fraenkel 84]).
def[OP: "TTT"; OP(x, y) = {{x}, {x, y}} ];    // Ordered Pair, instead of OP(x, y),
                                              // one can write just (x, y)
def[val: "TTT"; val(f, x) = H[y | (x, y) ∈ f ] ];  // the value of (the function) f at x
                                              // instead of val(f, x) one can write just (x, y)
def[F[: "DTT"; F[d, f] = {d, ([d], f)}];      // lambda-notation (it is a definition schema)

def[op: "TF"; op(z) ≡ E[x, y | z = (x, y) ] ];   // op(z) ≡ "z is an ordered pair"

def[rel: "TF"; rel(s) ≡ A[x:s, op(x)];           // rel(s) ≡ "s is a relation"

def[dom: "TT"; dom(f) = {x: U[f] | E[y|(x, y) ∈ f]} ]; // domain of the relation f

def[rng: "TT"; rng(f) = {y: U[U[f]] | E[x|(x, y) ∈ f]} ]; // range of the relation f

def[fn: "TF"; fn(f) ≡ rel(f) & A[x,y,y1 | (x, y) ∈ f & (x, y1) ∈ f → y = y1] ]; // fn(f) ≡
                                              // "f is a function"
def[oneonefn: "TF"; oneonefn(f) ≡ fn(f) & A[x,x1,y | (x, y) ∈ f & (x1, y) ∈ f → x = x1]];
                                              // oneonefn(f) ≡ "f is a one-one function
def[equivalent:"TTF"; equivalent(A, B) ≡ E[f | oneonefn(f) & dom(f) = A & rng(f) = B]];
                                              // equivalent(A, B) ≡ " the sets A and B are equivalent"

## 9. Another approach to defining relations and functions

We can follow [Bourbaki 68] and define a *relation* $\mathcal{R}$ as a triple <A, B, R>, where A, B are some sets (called the departure and arrival sets respectively) and the set R ⊆ A × B is called the graph of the relation. $\mathcal{R}$.
 In D-logic, this definition can be written in the following way:

$$\text{REL} := \text{def}[\text{Dep: set, Arr: set, R: P(Dep} \times \text{Arr) ]}$$

Let us now define some related concepts of the theory REL.

dom := {x: Dep & E[y:Arr, (x, y) ∈ R]}        // Domain of the relation

rng := {y: Arr & E[x:Dep, (x, y) ∈ R]}        // Range of the relation

func1 := [x:Dep ! y1:Arr ! y2:Arr, (x,y1) ∈ R & (x,y2) ∈ R → y1 = y2];
        // func1: the relation is functional on first argument
func2 := A[y:Arr ! x1:Dep ! x2:Dep, (x1,y) in R & (x2,y) in R → x1 = x2];
        // func2: the relation is functional on second argument
PFN := REL & func1;                           // Partial function

FN := PFN & Dep = dom;                        // Function

def[rel: "TTT"; rel(A, B) = { r:P(A, B), [A, B, r] } ];
        // rel(A, B) is the set of all relations on A×B
def[fn: "TTT"; fn(A, B) = {r: rel(A, B) & r.dom = r.Dep & r.func1}]d;
        // fn(A, B) is the set of all functions from A to B
def[ifn: "TTT"; ifn(A, B) = {f: fn(A, B) & f.func2} ];   // injective functions

def[sfn: "TTT"; sfn(A, B) = {f: fn(A, B) | f.rng = B} ]; // surjective functions

def[bfn: "TTT"; bfn(A, B) =  ifn(A, B) * sfn(A, B) ]d;   // bijective functions

def[equivalent: "TTT"; equivalent (A, B) ≡ bfn(A, B) ≠ { } ];

The definition of relation in this section can be called as "object-oriented", in contrast to the same definition in the previous section.

## 10 Conclusion

A new practical logical language, Predicate Logic with Definitions (or D-logic), based on first-order logic and also using some ideas of object-oriented programming has been described. Some preliminary experimentation with the D-logic has shown that D-logic can be useful both for practical formalization of mathematics and for the design of object-oriented specification languages.

## References

[Bourbaki 68]  Bourbaki N. Theory of sets. Addison-Wesley, 1968
[Davis 93]     Davis M. First Order Logic. In: Handbook of Logic in Artificial
        Intelligence and Logic Programming, vol 1. Clarendon Press, 1993.
[Dieudonne 82] Dieudonne J.A. A Panorama of Pure Mathematics.
        New York: Academic Press, 1982.
[Farmer 92]   Farmer W.M, Guttman J, D., Thayer F.J. Little theories.

      Lecture Notes in Computer Science, vol. 607, 1992, pp. 567-581.
[Fraenkel 84] Fraenkel A., Bar-Hillel Y., Levy A. Foundations of Set Theory.
      Amsterdam: Elsevier Science Publishers, 1984.
[Friedman 97] Friedman H. The Formalization of Mathematics.
      http://www.math.ohio-state.edu/foundations/ps/formofmath_5_21_97.ps
[Glushkov 72] V. M. Glushkov, Yu. V. Kapitonova, A.A. Letichevskiy et. al.
      Toward constructing a practical formal mathematical language
      for writing mathematical theories. Kibernetika, 1972, No. 5.
[Gordon 93] M.J.C. Gordon and T.F. Melham (eds). Introduction to HOL.
      Cambridge University Press: 1993
[Harrison 96] Harrison J. Formalized Mathematics. Technical Report 36, Turku Centre
      for Computer Science, also is available on the Web:
      http://www.cl.cam.ac.uk/users/jrh/papers/form-math3.html
[Hilbert 68] Hilbert D. und Bernays P. Grundlagen der Mathematik,
      Springer-Verlag, V. 1: 1968, V. 2: 1970.
[Jones 90] Systematic Software Construction using VDM. Prentice Hall, 1990.

[Kay 96] Kay A. The early history of SMALLTALK. In: History of programming
      languages, New York: ACM Press, 1996, pp. 511-598.
[Makarov 92] Makarov V. MSL - A Mathematical Specification Language.
      Lecture Notes in Computer Science, vol. 620, 1992, pp. 305-313.
[McAllister 88] McAllister D. Ontic: A Knowledge Representation System for
      Mathematics. The MIT Press, 1988.
[Mendelson 63] Mendelson E. Introduction to Mathematical Logic.
      D.Van Nostrand Company, Prinston: 1963.
[Meyer 92] Meyer B. Eifell: the Language. New York: Prentice Hall, 1992.

[QED Manifesto 94] The QED Manifesto. Lecture Notes in Computer Science,
      1994, vol. 814, pp. 238-251; also see: http://www.mcs.anl.gov/qed
[Sethi 96] Sethi, R. Programming languages: concepts & constructs, Addison-Wesley,
      1996.
[Shoenfield 73] Shoenfield J. Mathematical logic, Addison-Wesley, 1973.

[Spivey 92] Spivey J. The Z notation: A Reference Manual. Prentice Hall, 1992.

[Stroustrap 97] Stroustrap B. The C++ Programming Language. Addison-Wesley, 1997.

[Trybulec 93] Trybulec A. Some Features of the Mizar Language, ESPRIT Workshop,
      Torino, 1993. (also available at http://mizar.org/project/trybulec93.ps)
[Tseytin 98] Tseytin G. A Formalization of Reasoning not Derived from Standard
      Predicate Logic: http://www.math.spbu.ru/~tseytin/ARTICLE.ps.gz